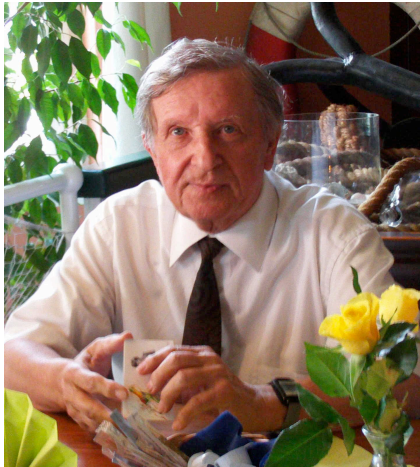
**Figure 1: Manfred Bonitz 2012, courtesy of Marc Bonitz**

Remembering Manfred Bonitz (7.3.1931 – 14.8.2012) on the first anniversary of his death

Devoted passionately to science and still full of ideas and plans, Manfred Bonitz, could not defeat his illness any longer and passed away last summer. Manfred Bonitz belongs to the pioneers in the information sciences in the socialist countries and devoted a substantial part of his scientific life to the newly emerging field of scientometrics. His membership in the editorial board of the journal *Scientometrics* lasted for more than 20 years. In this journal appeared most of his bibliometric publications, he celebrated winners of the Derek de Solla Price Medal in letters to the editor (Bonitz 1994), paid tribute to leading scientists in the field such as Robert K. Merton (Bonitz 2003), and acted as guest editor (Bonitz 2001).

We, one of his colleagues, and one of his sons (a scientist too) would like to take the opportunity of his first death anniversary to look back in greater detail to a unique scientific biography which mirrors main events in the course of the 20th century.

*Early years*

Manfred Bonitz was born in 1931 in Chemnitz, in South East Germany. His parents crafted furniture and had a small shop. His family, including his older sister, loved the bright and humorous boy, and he had enjoyed his childhood. His school years were massively influenced by Nazi propaganda until the regime was defeated in 1945. His final school years already took place under the rule of the Soviet forces in Germany. The dramatic switch of the system, reports and movies about Nazi cruelties, war crimes, and the suffering of Jews and the Soviet people determined his views for the rest of his life. He became an active anti-fascist and pacifist, and the friendship between Germany and Russia (and among all countries) was dear to his heart. East Germany, what in 1949 became the German Democratic Republic (GDR), appeared to Manfred a country that followed exactly these goals. After finishing High School in 1949, Manfred studied physics at the University of Jena and obtained his diploma degree in 1956. He then received an offer to continue his studies as a graduate student (then called aspirant) at the A.F. Ioffe Institute in Leningrad, in the Soviet Union, where he obtained his PhD in 1961. In the Soviet Union he found many friends who remained close to him for his whole life. Most importantly, of course, he found his great love, Natascha, whom he married in 1960 and whom he loved until her death in 2002. With Natascha and their first son, Michael, Manfred returned to East Germany (GDR) in 1961, they settled in Dresden which became the home for the family. His second son, Marc, was born there, in 1963. Many friends and colleagues visited him in his apartment with the beautiful view above the city and enjoyed the unique hospitality of Natascha and Manfred. In Dresden, Manfred worked under J. Schintlmeister, first at the Technical University and later at the Zentralinstitut für Kernforschung Rossendorf (ZfK, Central Institute for Nuclear Research, now the Helmholtz-Zentrum Dresden-Rossendorf) that remained his scientific home throughout his life.

*From physics to information science*

Manfred was a scientist with every fibre of his being. As said before, he started his scientific career in the – at that time – "hot" field of experimental nuclear physics in Jena under W. Schütz. At the A.F. Ioffe Institute in Leningrad he continued to work in nuclear physics under E.E. Berlovich. His main interest was to measure the lifetime of unstable nuclei, and he was very proud of his development of clever techniques that then set the world record for the shortest achievable time resolution (nanoseconds, Bonitz and Berlovich 1960, Bonitz 1963). At the ZfK Rossendorf he became the head of a small group and continued to measure lifetimes and excitations of nuclei. In 1965 he received an invitation to the famous Niels-Bohr Institute in Copenhagen where he worked until 1967 (e.g., Bonitz et al. 1968). He enjoyed the excellent working conditions and contacts to leading nuclear physicists. His nuclear physics research results were published in a total of about 30 papers and culminated in his habilitation (second doctoral) thesis in 1969 (Bonitz 1969). At the ZfK Rossendorf he had a reputation as good scientist, creative and humorous colleague (many recall his plays and sketches at various occasions) and also as a brilliant manager. At that time, shortage of all kinds of equipment was common, but Manfred often found a non-standard solution by exploiting his contacts with colleagues in different places.

In 1970 he was promoted to Head of a new department of the institute that was in charge of the library, data management, and the computing centre. In this position he, for the first time, faced the issues of how to access scientific information and how to efficiently organize this access on the level of a large institute. However, a few years later, following the death of his mentor, J. Schintlmeister, the new director, G. Flach, performed another reorganization of the resources after which Manfred eventually lost his position as department head. It was already too late to return to physics, so he decided to continue working in the young field of information science. Thus, he followed the same path as many physicists after World War II in the East and West - think, for example of Derek de Solla Price. This field was stimulated by the rapid rise of computers and automatic information processing; the fast growth of the entire science system, and the need of new scientific principles for information management. The step from nuclear physics to the information sciences was paved by the pioneering role nuclear physics has played in the emergence of the information sciences. Big research centres in the East and in the West very early on had to handle large amounts of experimental data and large numbers of scientific publications that had to be made rapidly accessible to all scientists. Even though Manfred loved and kept loving physics he started working in information science with great energy and enthusiasm. He established new contacts to information experts in Eastern Europe, most importantly in the Soviet Union, where this field did already have a longer tradition. Among these colleagues have been Mikhaïlov and Gilyarevskiĭ (Michajlov et al. 1980).

Although Manfred did not have a regular university appointment he loved teaching very much. So he used the chance to present lecture courses at various departments of the Technical University Dresden about modern information systems, cybernetics, and computer systems. His eight lectures for chemists were published as a ZfK report (Bonitz 1974). This lecture material was developed further into his excellent monograph *"Wissenschaftliche Forschung und wissenschaftliche Information [Scientific Research and Scientific Information]"*, published in German and in Russian (Bonitz 1979).

As a newcomer in the field of scientific information Manfred was not always met with open arms by the established communities that studied how science works. He, in contrast, came with his own experience in physics and with personal knowledge about the "real" information needs of a physicist. This brought many new ideas into the field and questioned many established points of view. He advocated modern information tools such as the *Science Citation Index* and journals such as *Scientometrics* and critically assessed the quality of scientific journals of the GDR (Bonitz 1983) generating herewith a lot of fuss in the science administration of the country. When he then openly advocated publishing in high impact (i.e. Western) journals he broke with another *Tabu* (taboo) and the fuss reached even the highest levels of the Academy of Sciences of the GDR (Bonitz 1985).

Besides publishing on information science, Manfred frequently was involved in communication of science to a broader public. This included writing reports (together with the director of the institute, G. Flach) on the progress of his institute that appeared in 1979, 1981, and 1986 (see, e.g., Flach and Bonitz 1986). Printed in a physics journal (*Kernenergie*) this publication stands as an example for his many publications *about* scientific communication and information in the field of his scientific origin. He also published reports on the peaceful use of nuclear energy in the GDR. After the nuclear disaster in Chernobyl where reliable information was difficult to obtain, he initiated a rapid, scientifically founded report (Bonitz 1988).

*Bibliography of Manfred Bonitz – a small scientometric excercise*
A bibliography of the publications of Manfred Bonitz  (Parthey 2003) listed his first information-science publication in the year 1970 with the title „The International Nuclear Information System INIS". (Bonitz 1970) The scientific work of Manfred Bonitz can be best illustrated relying on  a database which played an important part in his life – the *Science Citation Index* – now known as *Web of Science* or *Web of Knowledge*.
The 2001 bibliography of Manfred Bonitz lists 166 publications of very different nature: journal articles, popular science articles, monographs, and edited volumes, but also 21 reviews. He wrote in German, Russian, and English. The *Web of Science* contains 63 of his international publications (including editorials, letters to the editor, and book reviews). A closer inspection of this small selection already clearly marks three phases in his scientific career. [1]

---

[1]     For another, more extended bibliometric analysis of articles by Manfred Bonitz in the

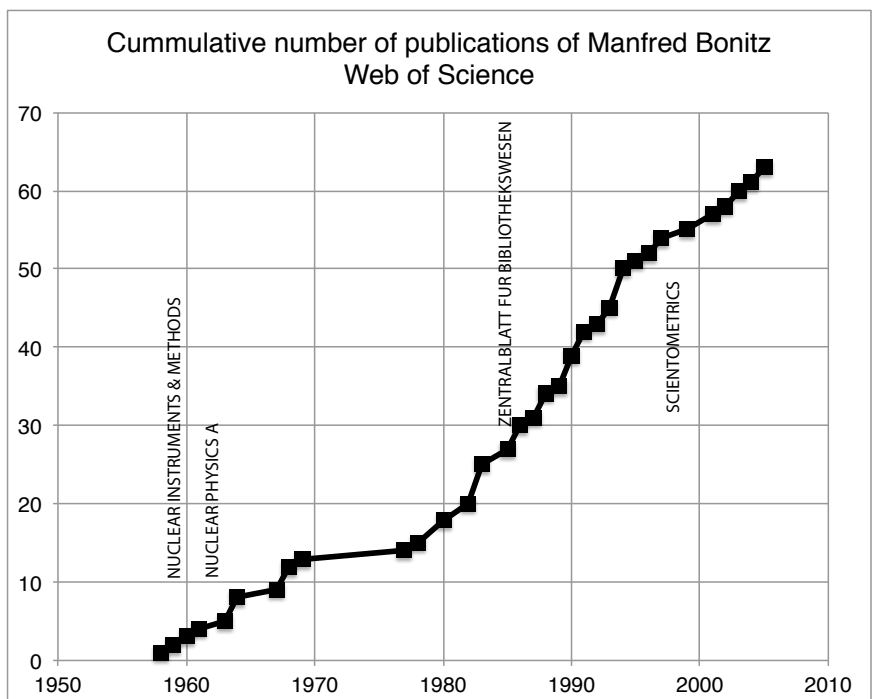

**Figure 2: Cumulative number of Manfred Bonitz's publications in the *Web of Science* and examples of journals**

In Figure 2, the following three periods in the career of Manfred Bonitz are clearly visible: from 1958 to the end of the 1960s, he published in physics journals, mostly co-authored articles (see above). The cumulative co-author network is shown in Figure 3. Between 1970 and 1980, the *Web of Science* records few international publications. His last pure physics publication in the *Web of Science* dates from 1969 (Funke et al. 1969), and the next publication listed there is from 1977 and clearly belongs to the information sciences (Bonitz 1977).

The left network in Figure 3 displays his collaboration network in the period 1958 to 1968 where many articles – as is typical for experimental nuclear physics – were multi-authored. The middle network shows how those collaborations in physics deepened and grew up to 1988, but also that some collaborations emerged in his new area of activity – the information sciences. A first collaborative article with P. Schmidt analyses communication in nuclear physics, but does so in an information science journal. (Bonitz and Schmidt 1978).

For the 1980s the *Web of Science* lists several book reviews about Garfield's "*Essays of an Information Scientist*" – work which inspired him and which he admired all his life. But he also writes in the German journal *Zentralblatt für Bibliothekswesen* (listed in the *Web of Science*!) about the newly emerging field of scientometrics, the Science Citation Index, the position of GDR publications in this international database; and, in parallel, in the journal *Scientometrics* about Bradford's law, journal rankings, and science-historically about Gustav Theodor Fechner, Wilhelm Ostwald, and the *Journal de Scavans*. With these publications he contributes to a diffusion of ideas between East and West in the area of information science.

---

*Social Science Citation Index* only see Leydesdorff (2011)..

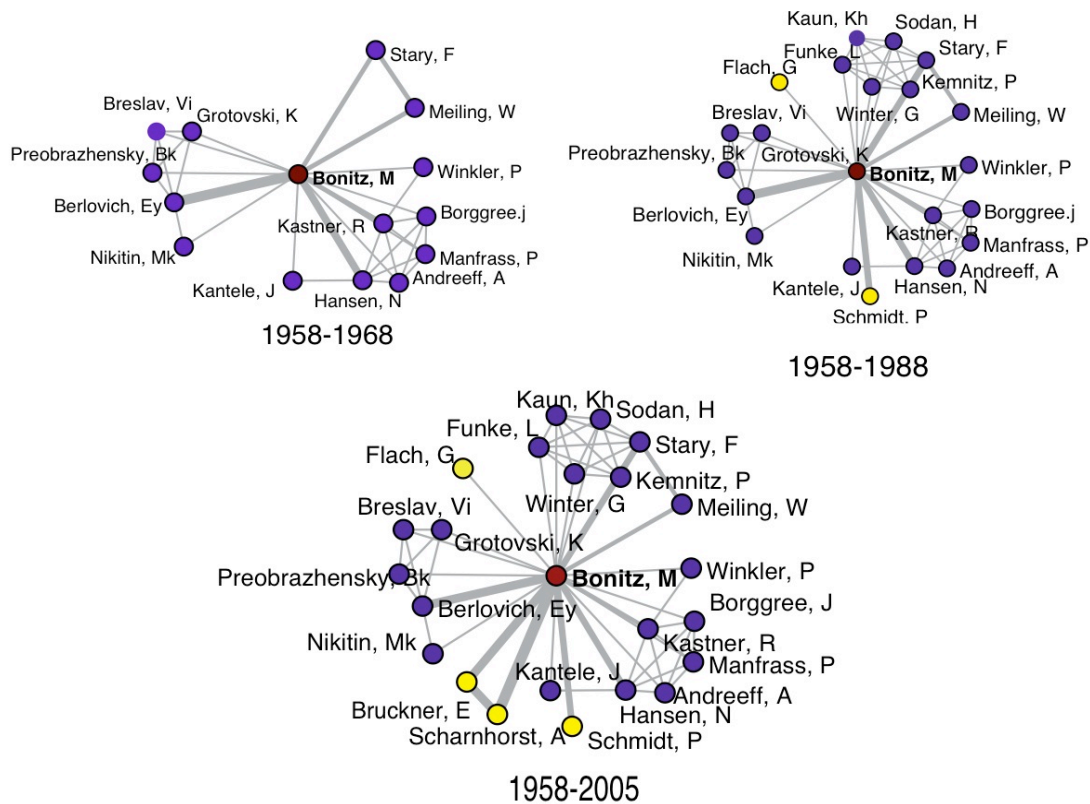

**Figure 3: Co-author network of Manfred Bonitz. The nodes are color-coded to the disciplines (physics=purple, social sciences=yellow).[2]**

In the early 1990s, the Academy of Sciences of the GDR was closed, and although the Central Institute for Nuclear Research survived and provided him with an affiliation address, job-wise he was expelled from the institute he had worked for all his life, a fate he shared with many other academics of the GDR. But, Manfred never lost his optimism and curiosity in scientific exploration and found new ways to continue research. And so the number of his international publications grew. Among the topics after the social transformation we find the 'Matthew effect of countries', but also science-historical work on Nalimov. Parallel to his international publications and his presence at international conferences, he remained active in the national science environment, continued also to participate in local and national German conferences in the information and documentation sciences.

*Manfred Bonitz's scientific heritage in the information sciences*
To characterize Manfred Bonitz's contribution to the information sciences, we would like to use a longer quotation of a text of Wolfgang Glänzel as a testimony. He wrote on the occassion of Manfred's 80th birthday: "[his] name is, beyond any doubt, most likely associated with studies of the Matthew Effect in scientometrics (Bonitz et al. 1997). Maybe the Matthew effect for countries, Journals and the Matthew Index are Manfred's best known results in this context. Less known is probably the fact that he has also introduced two basic principles of human behaviour in scientific research and communication. He formulated his principles as early as in the middle of the 1980s. And similarly to his idea of

---

applying the Mertonian notion of the Matthew Effect in the sciences (cf. Merton,1968[...]) to scientometrics, his universal behavioural principles governing research and communication processes are somewhat related to another basic principle of human behaviour, to Zipf 's Principle of Least Effort. This principle, which is only indirectly linked to Zipf 's empirical law, namely through the very property of human behaviour not to obey a Gaussian distribution but a rather to follow a power law, is a truly universal one as it can be observed in many areas like evolutionary biology, social behaviour or scholarly communication. [...] Manfred Bonitz has added two further principles to describe researchers' behaviour in seeking and disseminating information. He called his first law holography principle (cf. Bonitz, 1986) as it describes human behaviour as aiming at the broadest dissemination, access and retrieval of information. "Scientific information 'so behaves' that it is eventually stored everywhere. Scientists 'so behave' that they gain access to their information from everywhere." (Bonitz, 1991) Its 'temporal' counterpart, the maximum speed principle (cf. Bonitz, 1986) describes human behaviour as aiming at the fastest dissemination, access and retrieval of information. "Scientific information 'so behaves' that it reaches its destination in the shortest possible time. Scientists 'so behave' that they acquire their information in the shortest possible time." (Bonitz, 1991)

In fact, these two principles form one single law, in particular, they express the spatial-temporal duality of the same universal principle, the optimum dissemination, access and retrieval of information. By introducing this principle, Manfred proved a true visionary since little was known about the opportunities of the upcoming electronic communication at that time. The opportunities offered by the IT revolution, the electronic communication and the web have speeded up communication by several orders of magnitude. e-communication and e-publication, institutional and personal websites and repositories along with intelligent software solutions have increased visibility, extended storage capacities and facilitated access and retrieval of scientific information as never before. And scientists make an extensive use of these possibilities." (Glänzel, 2011)

*Epilogue*

In the canon of science, the information sciences are still a relatively young field, but one with already a considerable history of its own. Last year, ASIST – the former American Society for Information Science and Technology – devoted a whole day workshop to the celebration of the 75th anniversary of this remarkable professional society. (Carbo, Hahn 2012) However, continuously also new dimensions of the information sciences emerge, think for example of information visualization, semantic web technologies, and (inside scientometrics) altmetrics. To make contact with the history of our own field by looking into biographies of its early explorers is one way to understand and place current debates and controversies. As this account in honour of Manfred Bonitz shows, individual trajectories into and through science contain a manifold of important lessons. What has been the driving force for Manfred Bonitz all his life was scientific curiosity together with a deep humanity. Probably the most important lesson to be learned from Manfred's life is, that most important through all struggle of life is to try to be honest and responsible towards the scientific truth, but even more so towards your colleagues, friends, and family.

Michael Bonitz (Kiel)
Andrea Scharnhorst (Amsterdam, Berlin)